\documentclass[aps,prl,twocolumn,lengthcheck,groupedaddress,showpacs]{revtex4}
\usepackage{amsmath}
\usepackage{graphicx}
\usepackage{ifthen}
\usepackage{amsfonts}
\usepackage{amssymb}%
\def\cA{{\mathcal{A}}}

\def\bR{{\mathbb{R}}}

\def\vx{{\bf x}}
\def\vy{{\bf y}}

\def\be{\begin{equation}}
\def\ee{\end{equation}}
\newtheorem{theorem}{Theorem}
\newtheorem{corollary}[theorem]{Corollary}

\newtheorem{proposition}[theorem]{Proposition}

\begin{document}
\title{Predictability of extreme events in a branching diffusion model}

\author{Andrei Gabrielov} \affiliation{Departments of Mathematics and Earth and 
Atmospheric Sciences, Purdue University, West Lafayette, IN, 47907-1395}
\email{agabriel@math.purdue.edu}

\author{Vladimir Keilis-Borok} \affiliation{Institute of Geophysics and Planetary Physics,
and Department of Earth and Space Sciences, University of California Los Angeles, 
3845 Slichter Hall, Los Angeles, CA 90095-1567} 
\email{vkb@ess.ucla.edu}
 
\author{Ilya Zaliapin} \affiliation{Department of Mathematics and Statistics,
University of Nevada, Reno, NV 89557-0084, Corresponding author.}
\email{zal@unr.edu}
\date{\today}

\begin{abstract}
We propose a framework for studying predictability of extreme events 
in complex systems.
Major conceptual elements --- {\it direct cascading} or {\it fragmentation},
{\it spatial dynamics}, and {\it external driving} ---
are combined in a classical age-dependent multi-type branching diffusion 
process with immigration.
A complete analytic description of the size- and space-dependent 
distributions of particles is derived.
We then formulate an extreme event prediction problem and determine
characteristic
patterns of the system behavior as an extreme event approaches.
In particlular, our results imply specific premonitory deviations from 
self-similarity, which have been heuristically observed in real-world and 
modeled complex systems.
Our results suggest a simple universal mechanism of such premonitory patterns
and natural framework for their analytic study. 
\end{abstract}


\maketitle

\section{Introduction}
\label{intro}
{\it Extreme events} (also called critical transitions, disasters, 
catastrophes and crises) are a most important yet least understood 
feature of many natural and human-made processes.
Among examples are destructive earthquakes, El-Ni\~nos, 
economic depressions, stock-market crashes, and major terrorist acts.
Extreme events are relatively rare, and at the same time they inflict 
a lion's share of the damage to population, economy, and environment.
Accordingly, studying the extreme events is pivotal both for
fundamental predictive understanding of complex systems and for 
disaster preparedness (see \cite{KBS03,AJK05} and references therein).

In this paper we work within a framework that emphasizes mechanisms 
underlying formation of extreme events.
Prominent among such mechanisms is {\it direct cascading} or {\it fragmentation}.
Among other applications, this mechanism is at the heart of 
the study of 3D turbulence \cite{Fri96}. 
A statistical model of direct cascade is conveniently given by the
branching processes; they describe populations in which each 
individual can produce descendants (offsprings) according to some 
probability distribution.
A branching process may incorporate {\it spatial} dynamics,
several types of particles (multi-type processes), age-dependence
(random lifetimes of particles), and immigration due to external
driving forces \cite{AN04}.

In many real-world systems, observations are only possible within a specific 
domain of the phase space of a system.
Accordingly, we consider here a system with an {\it unobservable}
source of external driving ultimately responsible for extreme events.
We assume that observations can only be made on 
a {\it subspace} of the phase space. 
The direct cascade (branching) within a system starts with injection
of the largest particles into the source.
These particles are divided into smaller and
smaller ones, while spreading away from the source and eventually
reaching the subspace of observations. 
An important observer's goal is to locate the 
external driving source.
The distance between the observation subspace and the source thus becomes 
a natural control parameter.
An extreme event in this system can be defined as emergence
of a large particle in the observation subspace.
Clearly, as the source approaches the subspace of observation, 
the total number of observed particles increases, the bigger 
particles become relatively more frequent, and the probability 
of an extreme event increases. 
In this paper, we give a complete quantitative description of this 
phenomenon for an age-dependent multi-type branching diffusion process 
with immigration in $\mathbb{R}^n$.

It turns out that our model closely reproduces the major premonitory patterns 
of extreme events observed in hierarchical complex systems. 
Extreme events in such systems are preceded by transformation of size 
distribution in the permanent background activity 
(see {\it e.g.,} \cite{KBS03}). 
In particular, general activity increases, in favor of relatively strong 
although sub-extreme events. 
That was established first by analysis of multiple fracturing and seismicity 
\cite{KB96,RKB97}, 
and later generalized to socio-economic processes \cite{KSA05}.
Our results suggest a simple universal mechanism of such premonitory
patterns. 

\section{Model}
\label{model}
The system consists of particles indexed by their 
{\it generation} $k=0,1,\dots$.
Particles of zero generation ({\it immigrants}) are injected into the system 
by an external forcing. 
Particles of any generation $k>0$ are produced as a result of 
splitting of particles of generation $k-1$. 
Immigrants ($k=0$) are born at the origin 
${\bf x}:=(x_1,\dots,x_n) = {\bf 0}$ according to a 
homogeneous Poisson process with intensity $\mu$.
Each particle lives for some random time $\tau$ and then transforms (splits) into a 
random number $\beta$ of particles of the next generation.
The probability laws of the lifetime $\tau$ and branching $\beta$  
are rank-, time-, and space-independent.
New particles are born at the location of their 
parent at the moment of splitting.

The lifetime distribution is exponential:
$\mathsf{P}\{\tau<t\} = 1 - e^{-\lambda\,t},$ $\lambda>0$.
The conditional probability that a particle transforms into $n\ge 0$ 
new particles (0 means that it disappears) given that the transformation 
took place is denoted by $p_n$.  
The probability generating function for the number $\beta$ of
new particles is thus
\be
\label{branching_pgf}
h(s) = \sum_n p_n\,s^n.
\ee
The expected number of offsprings (also called the {\it branching number}) 
is $B:=E(\beta)=h'(1)$ (see {\it e.g.}, \cite{AN04}).

Each particle diffuses in $\mathbb{R}^n$ independently 
of other particles. 
This means that the density $p({\bf x,y},t)$ of a particle
that was born at instant $0$ at point ${\bf y}$ solves the equation
\be
\frac{\partial p}{\partial t}
= D\left(\sum_i \frac{\partial^2}{\partial x_i^2}\right)p
\equiv D\bigtriangleup_{\vx} p
\label{diffusion}
\ee
with the initial condition $p({\bf x,y},0) = \delta({\bf x-y})$.
The solution of (\ref{diffusion}) is given by
\be
p({\bf x,y},t)= \left(4\,\pi\,D\,t\right)^{-n/2}
\exp\left\{-\frac{|{\bf x-y}|^2}{4\,D\,t}\right\},
\label{sol}
\ee 
where $|{\bf x}|^2 = \sum_i x_i^2$.
%

\begin{figure}
\centering\includegraphics[width=.22\textwidth]{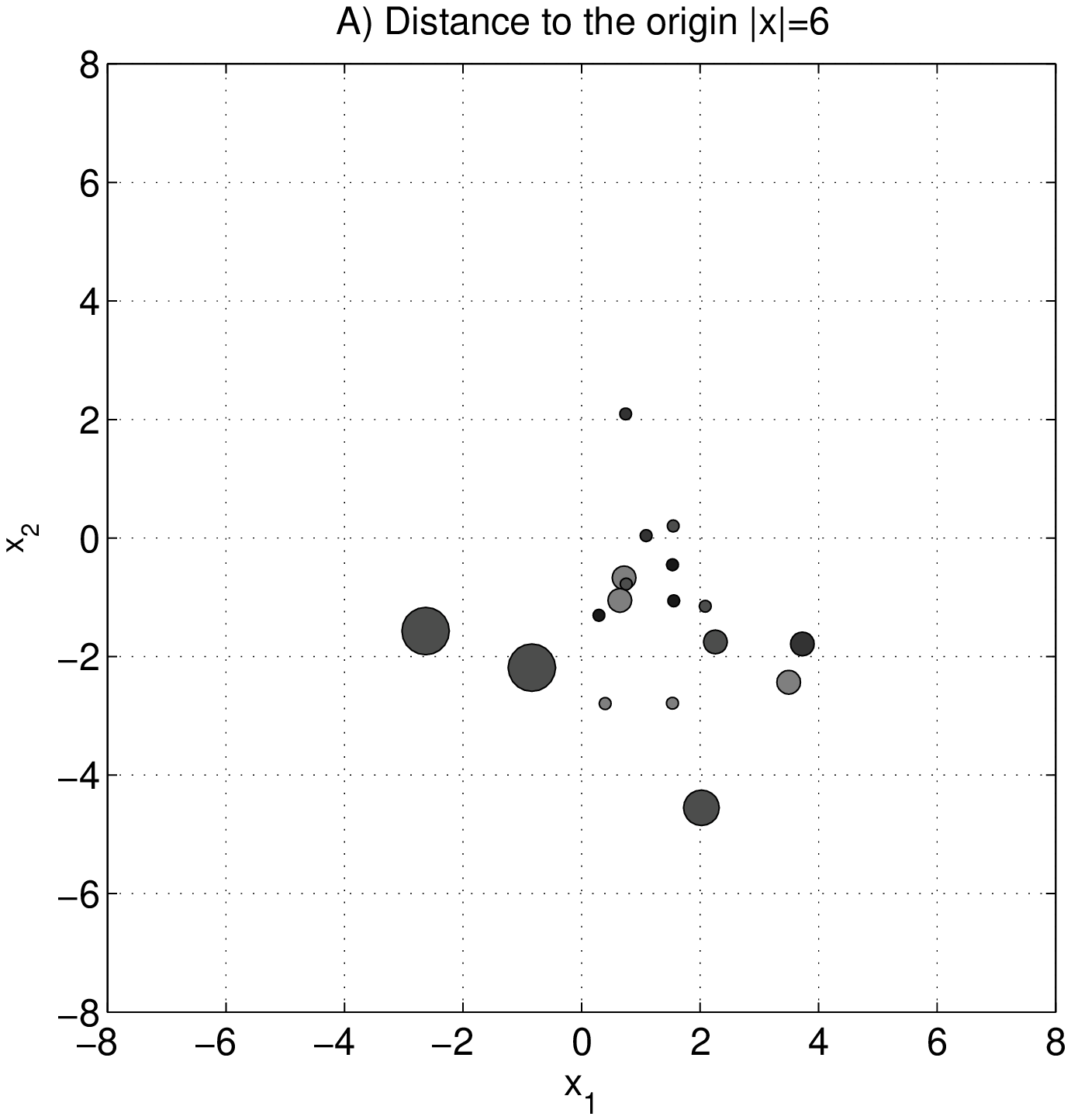}
\centering\includegraphics[width=.22\textwidth]{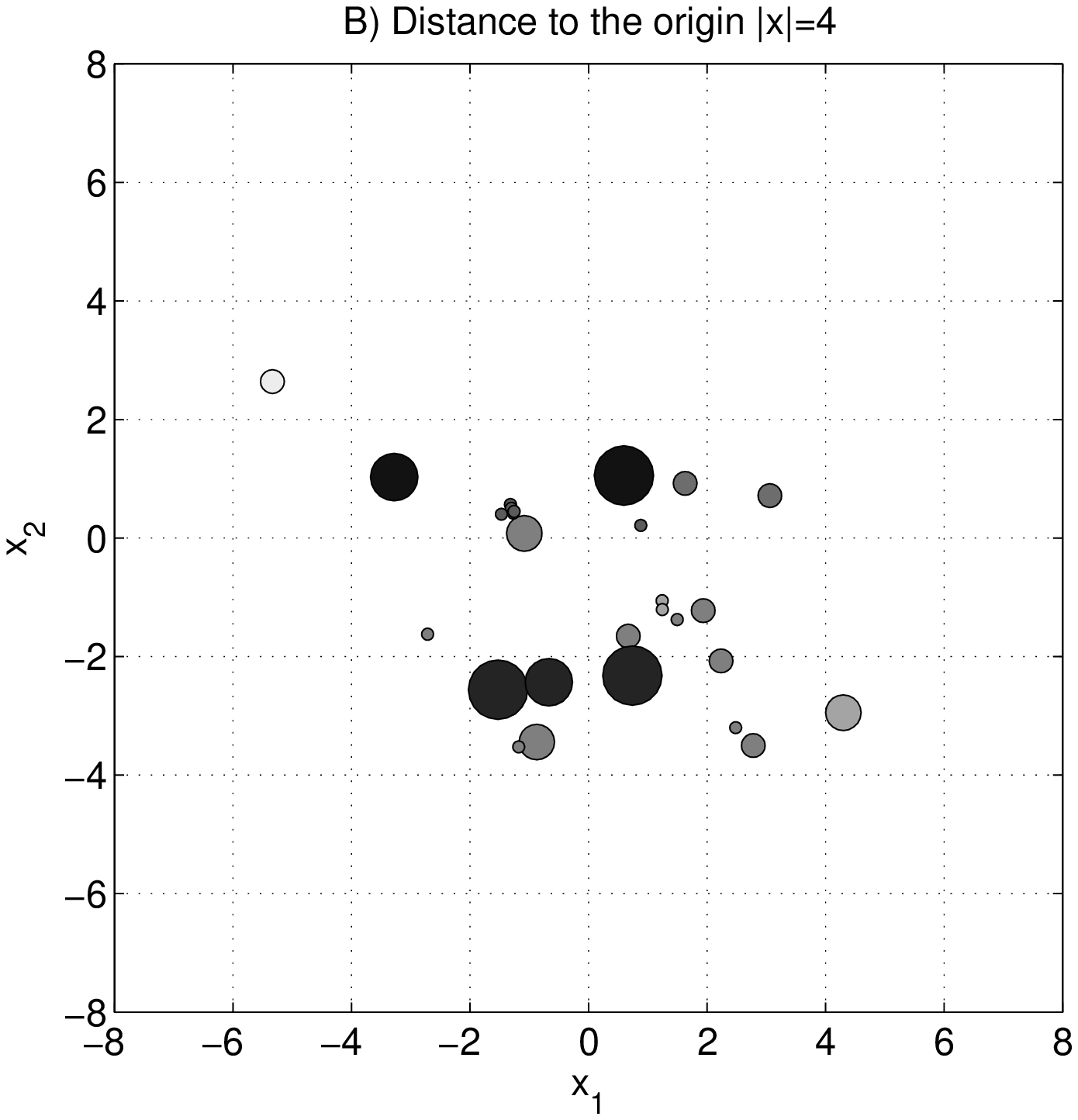}
\centering\includegraphics[width=.22\textwidth]{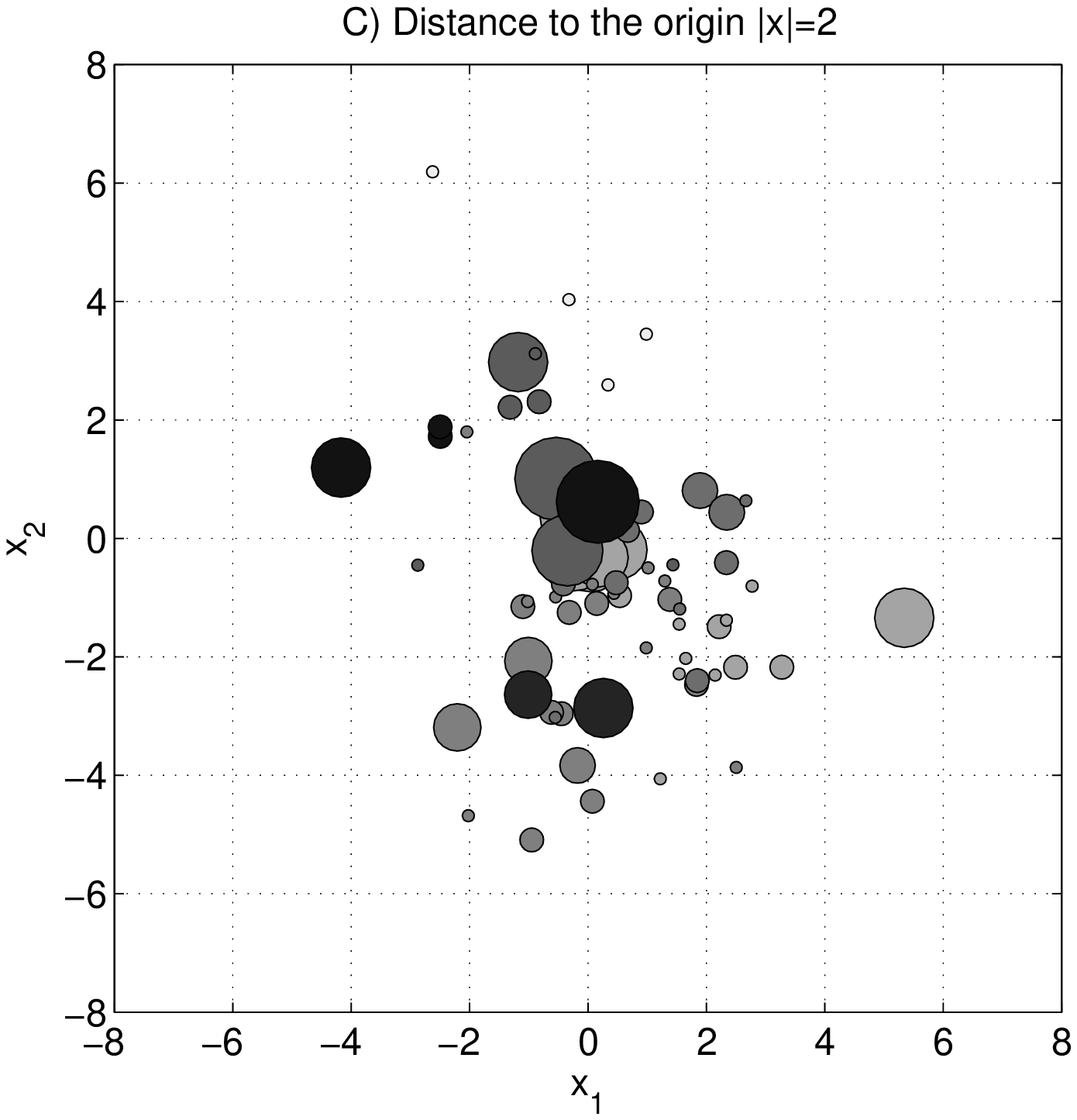}
\centering\includegraphics[width=.22\textwidth]{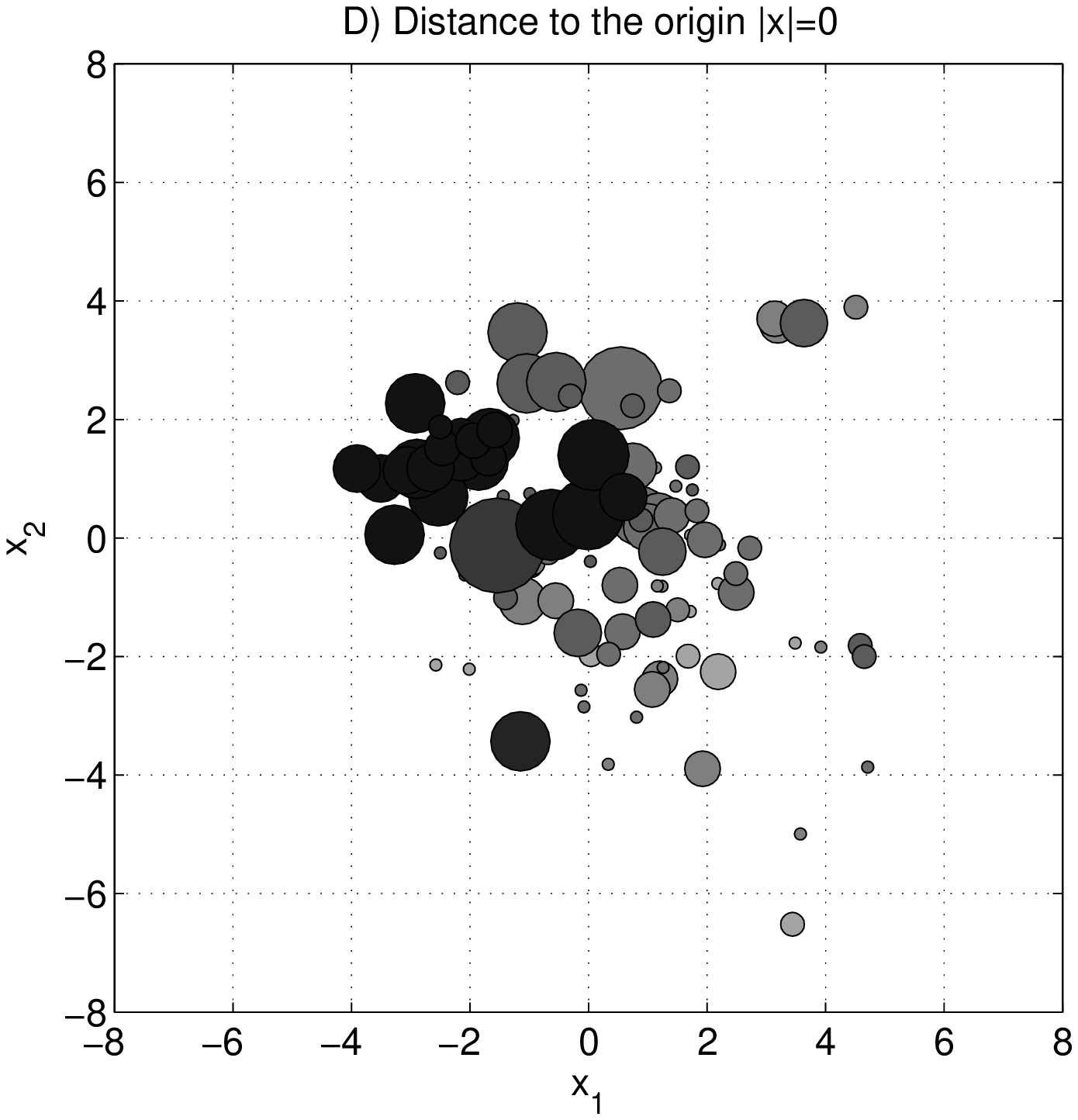}
\caption{Example of a 3D model population.
Different panels show 2D subspaces of the model 3D space
at different distances $|{\bf x}|$ to the origin. 
Model parameters are $\mu=\lambda=1$, $D=1$, $B=2$.
Circle size is proportional to the particle rank. 
Different shades correspond to populations from different 
immigrants, the descendants of earlier immigrants have lighter shade.
The clustering of particles is explained by the 
splitting histories.
Note that, as the origin approaches, the particle activity
significantly changes, indicating the increased probability
of an extreme event.
} 
\label{fig_example}
\end{figure}

It is convenient to introduce particle {\it rank} $r:=r_{\rm max}-k$ for an
arbitrary integer $r_{\rm max}$ and thus 
consider particles of ranks $r\le r_{\rm max}$. 
This reflects our focus on direct cascading, which often assumes
that particles with larger size  
({\it e.g.}, length, volume, mass, energy, momentum, {\it etc.}) split into 
smaller ones according to an appropriate conservation law.
Figure ~\ref{fig_example} illustrates the model dynamics. 

\section{Spatio-temporal particle rank distributions}
\label{results}

The model decsribed above is a superposition of independent branching 
processes generated by individual immigrants.
We consider first the case of a single immigrant;
then we expand these results to the case of multiple immigrants.
Finally, we analyze the rank distribution of particles.
Proofs of all statements will be published in a forthcoming paper.

\subsection{Single immigrant}
\label{single}

Let $p_{k,i}(G,{\bf y},t)$ be the conditional probability that at time $t\ge 0$
there exist $i\ge 0$ particles of generation $k\ge 0$ within spatial region 
$G\subset\mathbb{R}^n$ given that at time 0 a single 
immigrant was injected at point ${\bf y}$.  
The corresponding generating function is
\be
F_k(G,{\bf y},t;s) = \sum_{i}p_{k,i}(G,{\bf y},t)s^i.
\ee

\begin{proposition}
The generating functions $F_k(G,{\bf y},t;s)$ solve the following
recursive system of non-linear partial differential equations:
\be
\frac{\partial}{\partial t} F_k
=-D\bigtriangleup_{\bf y} F_k-\lambda\,F_k+\lambda\,h\left(F_{k-1}\right),\quad k\ge 1,
\label{pgf}
\ee
with the initial conditions $F_k(G,{\bf y},0;s)\equiv 1$, $k\ge 1$,
and
\be
F_0(G,{\bf y},t;s) = (1-P) + P\,s,
\label{pgf_ini}
\ee
where
$P:=e^{-\lambda\,t}\int_{G} p({\bf x,y},t)d{\bf x}.$
\label{prop_F}
\end{proposition}

Next, consider the expected number $\bar A_k(G,{\bf y},t)$ of
generation $k$ particles at instant $t$ within the region $G$
produced by a single immigrant injected at 
point ${\bf y}$ at time $t=0$.
It is given by the following partial derivative 
(see {\it e.g.}, \cite{AN04}) 
\be
\bar A_k(G,{\bf y},t):=\frac{\partial F_k(G,{\bf y},t;s)}{\partial s}\mid_{s=1}.
\ee
Consider also the expectation density $A_k({\bf x},{\bf y},t)$ that
satisfies, for any $G\subset\mathbb{R}^n$, 
\be
\bar A_k(G,{\bf y},t)=\int_G A_k({\bf x},{\bf y},t)\,d{\bf x}.
\label{gs}
\ee

\begin{corollary}
The expectation densities $A_k(\vx,\vy,t)$ solve the following recursive system
of linear partial differential equations:
\be
\frac{\partial A_k}{\partial t}
=D\bigtriangleup_{\bf x} A_k-\lambda\,A_k+\lambda\,B\,A_{k-1},\quad k\ge 1,
\label{exp_ave}
\ee
with the initial conditions $A_k(\vx,\vy,0)\equiv 0,~k\ge 1,$
\begin{eqnarray}
A_0({\bf x,y},0) &=& \delta({\bf y-x}),\nonumber\\
A_0({\bf x,y},t) &=& e^{-\lambda\,t} p({\bf x,y},t),~t>0.
\end{eqnarray}
The solution to this system is given by
\begin{eqnarray}
A_k({\bf x,y},t) = 
\frac{(\lambda\,B\,t)^k}{k!}A_0({\bf x,y},t).
\end{eqnarray}
\label{col}
\end{corollary}

The system \eqref{exp_ave} has a transparent intuitive meaning.
The rate of change of the expectation density $A_k({\bf x},{\bf y},t)$ is 
affected by the three processes: diffusion of the existing particles 
of generation $k$ in $\mathbb{R}^n$ (first term in the rhs of \eqref{exp_ave}), 
splitting of the existing particles of generation $k$ at the rate $\lambda$ (second term), 
and splitting of generation $k-1$ particles that produce on average $B$ new 
particles of generation $k$ (third term). 

\subsection{Multiple immigrants}
\label{multiple}

Here we expand the results of the previous section to the
case of multiple immigrants that appear at the origin 
according to a homogeneous Poisson process with intensity $\mu$.
The expectation $\cA_k$ of the number of particles of generation 
$k$ is given, according to the properties of expectations, by
\begin{eqnarray}
\cA_k({\bf x},t) = \int_0^t A_k({\bf x},{\bf 0},s)\,\mu\,ds
\end{eqnarray}
The steady-state spatial distribution $\cA_k({\bf x})$
corresponds to the limit $t\to \infty$ and is given by
\be
\cA_k(z)=\frac{\mu}
{\lambda\,k!}\left(\frac{B}{2}\right)^k
\left(\frac{2\,\pi\,D}{\lambda}\right)^{-n/2}\,
z^{\nu}\,K_{\nu}(z).
\label{Az}
\ee
Here $z:=|\vx|\sqrt{\lambda/D}$, $\nu=k-n/2+1$ and $K_{\nu}$ 
is the modified Bessel function of the second kind.

\subsection{Rank distribution and spatial deviations}
\label{GR}
Recall that the particle rank is defined as $r=r_{\rm max}-k$.
The spatially averaged steady-state rank distribution
is a pure exponential law with index $B$:
\begin{eqnarray}
A_k=\int\limits_{\bR^n}\int\limits_0^{\infty} 
A_k({\bf x,0},t) \mu\,dt\,d{\bf x}
=\frac{\mu}{\lambda}\,B^k\propto B^{-r}.
\label{pureexp}
\end{eqnarray}

To analyze deviations from the pure exponent, we
consider the ratio $\gamma_k({\bf x})$ between the number 
of particles of two consecutive generations:
\be
\gamma_k(\vx):=\frac{\cA_k(\vx)}{\cA_{k+1}(\vx)}.
\label{gamma}
\ee
For the purely exponential rank distribution, $A_k({\bf x}) = c\,B^k$, the 
value of $\gamma_k({\bf x})=1/B$ is independent of $k$ and ${\bf x}$; 
while deviations from the pure exponent will cause $\gamma_k$ to
vary as a function of $k$ and/or ${\bf x}$. 
Combining \eqref{Az} and \eqref{gamma} we find
\be
\gamma_k(\vx)=\frac{2\,(k+1)}{B\,z}\,
\frac{K_{\nu}(z)}{K_{\nu+1}(z)},
\label{gamma1}
\ee
where, as before, $z:=|\vx|\,\sqrt{\lambda/D}$ and $\nu=k-n/2+1$.

\begin{proposition}
The asymptotic behavior of the function $\gamma_k(z)$ is given by
\begin{eqnarray}
\lim\limits_{z\to 0}\gamma_k(z)&=&\left\{
\begin{array}{cc}
\infty,&\nu\le 0,\\
\displaystyle\frac{1}{B}\left(1+\frac{n}{2\,\nu}\right),&
\nu>0,
\end{array}\right.
\label{gammaz0}\\
\gamma_k(z)&\sim&
\frac{2(k+1)}{B\,z},~{z\to\infty},~{\rm fixed~}k,
\label{gammazinf}\\
\gamma_k(z)&\sim&
\frac{1}{B}\left(1+\frac{n}{2\,\nu}\right),~{k\to\infty},
~{\rm fixed~}z.
\label{gammakinf}
\end{eqnarray}
\label{gammalim}
\end{proposition}

\begin{figure}
\centering\includegraphics[width=.45\textwidth]{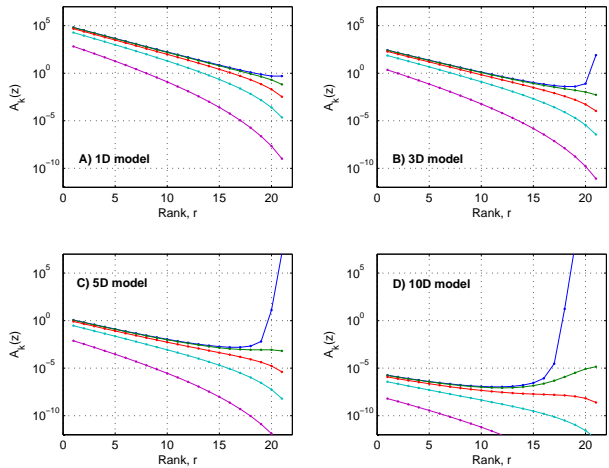}
\caption{Deviations from self-similarity:
Expected number $A_k(z)$ of generation $k$ particles 
at distance $z$ from the origin (cf. Proposition \ref{gammalim}). 
The distance $z$ is increasing
(from top to bottom line in each panel) as $z=10^{-3},2,5,10,20$.
Model dimension is $n=1$ (panel A), $n=3$ (panel B),
$n=5$ (panel C), and $n=10$ (panel D).  
Other model parameters: $\mu=\lambda=1$, $D=1$, $B=2$, $r_{\rm max}=21$. 
One can clearly see the transition from downward to upward
deviation of the rank distributions from the pure exponential form as 
we approach the origin.}
\label{fig_Ak}
\end{figure}

Proposition \ref{gammalim} allows one to describe all
deviations of the particle rank distribution from the pure
exponential law \eqref{pureexp}.
Figure~\ref{fig_Ak} illustrates our findings.
First, Eq.~\eqref{gammakinf} implies that at any spatial point, the 
distribution asymptotically approaches the exponential form as 
rank $r$ decreases (generation $k$ increases).
Thus the deviations can only be observed at the largest ranks 
(small generation numbers).
Analysis of the large-rank distribution is done using 
Eqs.~\eqref{gammaz0},\eqref{gammazinf}. 
Near the origin, where the immigrants enter the system,
Eq.~\eqref{gammaz0} implies that 
$\gamma_k(z) > \gamma_{k+1}(z)>1/B$ for $\nu >0$.
Hence, one observes the {\it upward deviations} from the pure exponent:
for the same number of rank $r$ particles, the number of
rank $r+1$ particles is larger than predicted by \eqref{pureexp}.
The same behavior is in fact observed for $\nu\le 0$ 
(the details will be published elsewhere).
In addition, for $\nu\le 0$ the ratios $\gamma_k(z)$ do not merely
deviate from $1/B$, but diverge to infinity at the origin.
Away from the origin, according to Eq.~\eqref{gammazinf}, we have 
$\gamma_k(z)<\gamma_{k+1}(z)<1/B$, which implies {\it downward deviations}
from the pure exponent: for the same number of rank $r$ particles,
the number of rank $r+1$ particles is smaller than predicted by \eqref{pureexp}.
%

%
%

\section{Discussion}
Motivation for this work is the problem of prediction of extreme 
events in complex systems.
Our point of departure is a classical model 
of spatially distributed population of particles of different 
ranks governed by direct cascade of branching and external driving.
In the probability theory this model is known as the age-dependent
multi-type branching diffusion process with immigration \cite{AN04}.
We introduce here a new approach to the study of this process.
We assume that observations are only possible on a subspace of 
the system phase space while the source of external driving 
remains unobservable.
The natural question under this approach is the dependence
of size-distributions of particles on the distance to the source.
The complete analytical solution to this problem is
given by the Proposition~\ref{prop_F}.

It is natural to consider rank as a logarithmic 
measure of the particle size.
If we assume a size-conservation law in the model,
the exponential rank distriburtion derived in
\eqref{pureexp} corresponds to a self-similar, power-law 
distribution of particle sizes, characteristic for many 
complex systems.
Thus, the Proposition~\ref{gammalim} describes space-dependent 
deviations from the self-similarity (see also Fig.~\ref{fig_Ak});
in particular, deviations premonitory to an extreme event.  
The numerical experiments (that will be published elsewhere) 
confirm the validity of our analytical results and asymptotics 
in a finite model.

The model studied here exhibits very rich and intriguing 
premonitory behavior. 
Figure~\ref{fig_example} shows several 2D snapshots of a 3D model
at different distances from the source.
One can see that, as the source approaches, the following
changes in the background activity emerge:
a) The intensity (total number of particles) increases;
b) Particles of larger size become relatively more numerous;
c) Particle clustering becomes more prominent;
d) The correlation radius increases;
e) Coherent structures emerge.
In other words, the model exhibits a broad set of premonitory phenomena
previously observed heuristically in real and modeled systems: 
multiple fracturing \cite{RKB97}, 
seismicity \cite{KB96},
socio-economics \cite{KSA05},
percolation \cite{ZWG04},
hydrodynamics, 
hierarchical models of extreme event development \cite{KBS03}.
These phenomena are at the heart of earthquake prediction
algorithms well validated during 20 years of forward world-wide 
tests (see {\it e.g.,} \cite{KBS03}).

In this paper we analyse only the first-moment properties of
the system; such properties can explain the premonitory 
intensity increase (item a above) and transformation of the particle 
rank distribution (item b).
At the same time, the framework developed here allows
one to quantitatively analyze other premonitory phenomena; 
this can be readily done by considering the higher-moment properties. 

\acknowledgments
This study was supported in part by NSF grant No. ATM 0620838.

\end{document}